\documentstyle[12pt]{article}
\def\beq{\begin{equation}}
\def\eeq{\end{equation}}
\def\bea{\begin{eqnarray}}
\def\eea{\end{eqnarray}}
\def\D{\Delta}
\def\b{\beta}
\def\a{\alpha}
\begin {document}
\begin{titlepage}
\begin{flushright}
HU Berlin-IEP-95/1
\end{flushright}
\mbox{ }  \hfill hepth@xxx/9501019
\vspace{6ex}
\Large
\begin {center}
\bf{Some Conclusions for Noncritical
String Theory Drawn from Two- and Three-point Functions in the Liouville
Sector}
\footnote{to
appear in the proceedings of the \em XXVIII. Int. Symp. on the Theory of
Elementary Particles \\
\hspace*{6cm} Wendisch-Rietz, August 30 - September 3, 1994 \em}
\end {center}
\large
\vspace{3ex}
\begin{center}
H. Dorn and H.-J. Otto \footnote{e-mail: dorn@ifh.de , otto@ifh.de}
\end{center} \normalsize
\it \vspace{3ex}
\begin{center} Humboldt--Universit\"at zu Berlin \\
Institut f\"ur Physik, Theorie der Elementarteilchen \\ Invalidenstra\ss e 110,
D-10115 Berlin \end{center}
\vspace{6ex} \rm
\begin{center} \bf{Abstract}
\end{center}
\vspace{3ex}
Starting from the known expression for the three-point correlation functions
for Liouville exponentials with generic real coefficients
at we can prove the
Liouville equation of motion at the level of three-point functions.
Based on the
analytical structure of the correlation functions we discuss a possible
mass shell condition for excitations of noncritical strings and make some
observations
concerning correlators of Liouville fields.
\end {titlepage} \newpage \setcounter{page}{1}
\pagestyle{plain}
{\Large \bf Introduction}\\ \\
One of the outstanding unsolved problems in string theory is posed by
the question how to go off criticality without taking refuge to Kaluza-Klein
concepts or saturating the conformal anomaly in D=4 by some additional degrees
of freedom. After Polyakovs seminal paper \cite{Pol} we believe to know
that the key lies in solving the 2D quantized Liouville field theory. Though
there has been a lot of progress on this subject in the course of the years
(\cite{Gervais1,Gervais2,QG,OW,KN,GSchn} and references therein )
we still do not know how to enter
the most interesting region $1<c_M<25$ for the Virasoro central charge $c_M$
of the matter system. In the bosonic string theory $c_M$ can be identified
with the embedding dimension $d$. A lot of interesting results especially
in comparison with solvable matrix-models were obtained for noncritical strings
for $c_M\leq 1$ (e.g. \cite{FGZ} and references contained), especially
at $d=1$ the noncritical string could be
treated \cite{GL,FK} and effectively resulted in a $D=1+1$-dimensional
 critical string due to
the extra dimension defined by the Liouville field.
In a previous paper \cite{DO2} we were able to generalize the calculation
of three-point functions to $d>1$ to become independent of the special
1d-kinematics. As we unfortunately still have to keep $c_M<1$ we have
to introduce background charges in target space sacrificing Lorentz
invariance or its euclidean counterpart. Here we are going to draw some
conclusions based on these results.

The central task in solving the theory of conformal matter coupled
to 2D gravity consists in calculating the correlators for all marginal
"dressed" operators
\footnote{We explicitely write only the "holomorphic" variable $z$ though
both $z$ and $\bar{z}$ are present in the arguments.}
\beq
\Omega _i^{(M)} e^{\beta _i \phi (z)}=\Omega _i^{(dressed)}
\label{e1}
\eeq
with conformal weights $\D _i^{(M)}$ and $\D _i^{(L)}$ in the matter and
Liouville sector, respectively, satisfying
\beq
\D _i^{(M)} + \D ^{(L)}(\b _i) = 1 .
\label{e2}
\eeq
This is the motivation to look at correlators of products purely of Liouville
exponentials for $N\geq 3$ :
\beq
G_{N}(z_{1},...,z_{N}\vert \beta_{1},...,\beta_{N})~=~ \langle \prod _{j=1}
^{N}e^{\beta_{j}\phi (z_{j})} \rangle
{}~=~ \int D\phi ~e^{-S_{L}[\phi \vert \hat{g}]} \prod _{j=1}^{N}
e^{\beta_{j} \phi (z_{j})}
\label{1}
\eeq
with generic real dressing coefficients $\b _i$ and the Liouville action
\beq
S_{L}[\phi \vert \hat{g}]~=~\frac{1}{8\pi} \int d^{2}z \sqrt{\hat{g}}
\Big(\hat{g}^{mn}\partial_{m}
\phi \partial_{n}\phi~+~Q\hat{R}(z)\phi(z)~+~\mu^{2}e^{\alpha \phi (z)}\Big)~.
\label{2}
\eeq
$\hat{g}$ is a classical reference metric of the 2D manifold of spherical
topology, $\hat{R}$ the corresponding Ricci scalar. $Q$ parametrizes the
central charge $c_{L}$ of the Liouville theory by
\beq
c_{L}~=~1~+~3Q^{2}~.
\label{3}
\eeq
As the Liouville theory describes the gravitational sector of a conformal
matter theory the conformal anomalies add up to zero
$c_{L}~+~c_{M}~-~26~=~0~$.
The exponent $\a $ in the cosmological term in (\ref{2})
is just the dressing coefficient for the unity operator in the matter sector
and derives from (\ref{e2}) for $\D ^{(M)}=1 ,~~\b _i \rightarrow \a$ and
the general formula for the conformal weight of a vertex operator
$exp(\b _i\phi )$ with a background charge $Q$ present in (\ref{2})
\beq
\D _i^{(L)}\equiv \Delta_{i}~=~\frac{1}{2} \beta _{i}(Q~-~\beta_{i})~.
\label{10}
\eeq
With the additional demand to the cosmological operator to be a
``microscopic" operator, i. e. $\alpha <\frac{Q}{2}$ \cite{Seiberg} ,
one gets
\beq
\alpha~=~\alpha_{-}~~~~~~~~
\alpha_{\pm}~=~\frac{Q}{2}~\pm ~\frac{\sqrt{Q^{2}-8}}{2}~.
\label{6}
\eeq
The zero mode integration in eq.(\ref{1})
can be performed explicitly \cite{GL}.
For integer
\beq
s_{N}~=~\frac{Q-\sum_{j=1}^{N}\beta_{j}}{\alpha} \label{7}
\eeq
also the remaining functional integral can be done:
\bea G_{N}(z_{1},...,z_{N}
\vert \beta_{1},...,\beta_{N})~=~\frac{\Gamma (-s_{N})}
{\alpha} \Big( \frac{\mu
^{2}}{8\pi} \Big) ^{s_{N}}~\prod _{1\leq i<j\leq N} \vert z_{i}~-~z_{j} \vert
^{-2\beta_{i} \beta_{j}} \nonumber \\ \cdot \int \prod _{I=1}^{s_{N}} \Big(
d^{2}w_{I} \prod _{j=1}^{N}\vert z_{j}~-~ w_{I}\vert ^{-2\alpha \beta_{j}}\Big)
\prod _{1\leq I<J\leq s_{N}} \vert w_{I}~-~w_{J}\vert ^{-2\alpha^{2}}~.
\label{8}
\eea

Let us start with the 3-point function.
Fortunately from the explicit representation (\ref{8}) we
can prove for integer $s_{3}$ that the standard structure of the $z_{j}$
dependence is realized.
\bea
G_{3}(z_{1},z_{2},z_{3} \vert \beta _{1},\beta_{2},\beta_{3})&=&
A_{3}(\beta_{1},\beta_{2},\beta_{3})~
\vert z_{1}-z_{2} \vert ^{2(\Delta _{3}-\Delta _{1}-\Delta _{2})}
\nonumber \\
&\cdot &\vert z_{1}-z_{3} \vert ^{2(\Delta _{2}-\Delta _{1}-\Delta _{3})}
\vert z_{2}-z_{3} \vert ^{2(\Delta _{1}-\Delta _{2}-\Delta _{3})}~,
\label{11}
\eea
\beq
A_{3}~=~\lim_{u_{3}\to \infty}\vert u_{3}\vert ^{4\Delta _{3}}G_{3}
(0,1,u_{3}\vert \beta_{1},\beta_{2},\beta_{3})~.
\label{12}
\eeq
With (\ref{8}) this gives $A_{3}$ as
\beq
A_{3}~=~\frac{\Gamma(-s_{3})}{\alpha} \Big ( \frac{\mu ^{2}}{8\pi}\Big )
^{s_{3}}\int \prod _{I=1}^{s_{3}}(d^{2}w_{I}\vert w_{I}\vert ^{-2\alpha \beta _
{1}}\vert 1-w_{I} \vert ^{-2\alpha \beta _{2}})
\prod _{1\leq I<J\leq s_{3}}\vert w_{I}-w_{J} \vert ^{-2\alpha ^{2}}~.
\label{13}
\eeq
Using the Dotsenko-Fateev integrals \cite{DF} this can be written as \cite{DO2}
\beq
A_{3}(\beta _{1},\beta _{2},\beta _{3})~=~
\frac{\Gamma (-s_{3})}{\alpha}~\Gamma (1+s_{3})\Big (\frac{\mu ^{2}~ \Gamma
(1+\frac{\alpha ^{2}}{2})}{8~\Gamma(-\frac{\alpha ^{2}}{2})}\Big ) ^{s_{3}}
\prod _{i=0}^{3}F_{i}
\label{14}
\eeq
with
\beq
F_{i}~=~\exp \Big (f(\alpha \bar{\beta _{i}},\frac{\alpha ^{2}}{2} \vert
s_{3})-
f(\alpha \beta _{i}, \frac{\alpha ^{2}}{2}\vert s_{3})\Big )~,~~~~~~~~~~
i=1,2,3
\label{15}
\eeq
\beq
\bar{\beta _{i}}~=~\frac{1}{2}(\beta _{j}+\beta _{k}-\beta _{i})~
=~\frac{1}{2}(Q-\alpha s_{3})-\beta _{i}~,
{}~~~~~~(i,j,k)=\mbox{perm}(1,2,3)~,
\label{16}
\eeq
\beq
F_{0}~=~\Big (-\frac{\alpha ^{2}}{2}\Big ) ^{-s_{3}}\frac{1}{\Gamma (1+s_{3})}
\exp \Big ( f(1-\frac{\alpha ^{2}}{2}s_{3},\frac{\alpha ^{2}}{2} \vert s_{3})
-f(1+\frac{\alpha ^{2}}{2},\frac{\alpha ^{2}}{2} \vert s_{3})\Big )~,
\label{17}
\eeq
\beq
f(a,b\vert s)~=~\sum_{j=0}^{s-1} \mbox{log}~ \Gamma (a+bj)~,~~~~~\mbox{integer}
{}~s~.
\label{18}
\eeq
There exists a continuation \cite{DO2,DO3,DO4}
of $f(a,b\vert s)$ to arbitrary complex
$a,b,s$ given by
\bea
f(a,b\vert s)=\int_{0}^{\infty }\frac{dt}{t}\Big(s(a-1)
e^{-t}+b\frac{s(s-1)}{2}
e^{-t}-s\frac{e^{-t}}{1-e^{-t}} \nonumber\\
+\frac{(1-e^{-tbs})e^{-at}}{(1-e^{-tb}) (1-e^{-t})}\Big).
\label{23}
\eea
It fulfills all the functional relations
that can be read off the representation
(\ref{18}) for integer $s$ .  Using the integral representation and the
functional relations mentioned
one can prove \cite{DO2} that exp$(f(a,b\vert s))$ is a meromorphic function.
It is sufficient to
investigate the case Re $b\geq 0$. Under this circumstance exp$f$ has poles at

\beq
a~~~=~-bj~-~l~~~~~~~~~~\mbox{(poles)}
\label{24}
\eeq
and zeros at
\beq
a~+~bs~=~-bj~-~l~~~~~~~~~~\mbox{(zeros)}~.
\label{25} \eeq
In both cases $j$
and $l$ are integers $\geq 0$.
In this talk we are going to focus on four applications of the foregoing
results :\\ \\
\noindent
(i) Correlator for two Liouville exponentials\\
(ii) Validity of the quantum Liouville equation\\
(iii) Two- and three-point function for the Liouville field itself\\
(iv) Conclusions about the poles and zeros of $A_2, A_3$ \\ \\
{\Large \bf (i) Correlator for two Liouville exponentials}\\ \\
Let us now turn to the 2-point function. Taking (\ref{1}) unmodified also for
$N=2$ would imply
$G_{2}(z_{1},z_{2}\vert \beta _{1},\beta _{2})~=~
G_{3}(z_{1},z_{2},z_{3}\vert \beta _{1},\beta _{2},0)$.
The unwanted
$z_{3}$-dependence as usual in conformal theories drops for $\Delta _{1}=
\Delta _{2}$. However, the $z$-independent factor $A_{3}(\beta ,\beta , \beta
_{3})$ diverges for $\beta _{3} \rightarrow 0$.
The reason for this
divergence is the change of the situation with respect to the conformal Killing
vectors (CKV). The 3-punctured sphere has no CKV's while the 2-punctured
sphere has one. The (divergent) volume of the corresponding subgroup
of the M\"obius group SL(2,C) leaving $z_{1}$ and $z_{2}$ fixed is
\beq
V^{(2)}_{CKV}~=~\int \frac{d^{2}w~\vert z_{1}-z_{2}\vert ^{2}}
{\vert z_{1}-w\vert ^{2}~\vert z_{2}-w\vert ^{2}}~~.
\label{27}
\eeq
Having this in mind we define
\beq
G_{2}(z_{1},z_{2}\vert \beta )~=~\langle e^{\beta \phi (z_{1})}e^{\beta \phi
(z_{2})}\rangle ~=~\frac{1}{V^{(2)}_{CKV}}\int D\phi ~e^{-S_{L}[\phi ]}
e^{\beta \phi (z_{1})}e^{\beta \phi (z_{2})}~~.
\label{28}
\eeq
Treating the functional integral in analogy to that for the 3-point function
and
choosing $\int d^{2}w_{1}$ as the cancelled integration one gets
$(s_2=1+s_3(\b ,\b ,\a ))$
\beq
G_{2}(z_{1},z_{2}\vert \beta)~=~-\frac{\mu ^{2}}{8\pi
s_{2}}G_{3}(z_{1},z_{2},w_{1}\vert \beta ,\beta ,\alpha)~\vert z_{1}-z_{2}
\vert
^{-2}~\vert z_{1}-w_{1}\vert ^{2}~\vert z_{2}-w_{1}\vert ^{2}~.
\label{30}
\eeq
{}From (\ref{11}) we see that the $w_{1}$ dependence on the r.h.s. cancels.
For this result $\Delta _{1}=\Delta _{2}$ is crucial. Altogether we find
\beq
G_{2}(z_{1},z_{2}\vert \beta )~=~\frac{A_{2}(\beta )}{\vert z_{1}-z_{2}\vert
^{4\Delta}}~~,
\label{31}
\eeq
with
\beq
A_{2}(\beta )~=~-\frac{\mu ^{2}}{8\pi s_{2}}~A_{3}(\beta ,\beta ,\alpha )~~.
\label{32}
\eeq
For this constellation of arguments in $A_3$ one can eliminate
the function $f$
completely and derive quite a simple expression in terms
of $\Gamma$-functions \cite{DO4}
coinciding up to an irrelevant factor
with the result presented in \cite{GL} for the integrated
2-point function in gravitationally dressed minimal models (rational
$s_{2}$). For $A_{2}$ describing the gravitational dressing of the two point
function in minimal models the resuling form fits into the ``leg-factor"
structure known for higher correlation functions ($N\geq 3$), \cite{FK}.
The extension of the procedure to the one and zero-point function (partition
function) is straightforward \cite{DO4} in principle.
However, in the one-point case the method yields inconsistent
results, the dependence on the fixed unintegrated $w_{1},~w_{2}$ does not
cancel for generic $\beta $.
This is a reflection of the absence of a SL(2,C) invariant vacuum,
which prevents looking at the one-point function as a scalar product of two
physical states. We come back to this point later.\\ \\ \\
%
%
%
%
%
{\Large \bf (ii) Validity of the Liouville Equation}\\ \\

The Liouville equation in our parametrization is the equation of motion for the
action (\ref{2}) in the limit of flat $\hat{g}$
\beq
\partial ^{2}~\phi ~-~\frac{\alpha \mu ^{2}}{2}~e^{\alpha \phi}~=~0~.
\label{36}
\eeq
As a partial check we want to prove
\beq
\langle \partial ^{2}\phi (z_{1})~e^{\beta _{2}\phi (z_{2})}~
e^{\beta _{3}\phi (z_{3})}\rangle~=~\frac{\alpha \mu ^{2}}{2}\langle
e^{\alpha \phi (z_{1})}~e^{\beta _{2}\phi (z_{2})}~e^{\beta _{3}\phi (z_{3})}
\rangle
\label{37}
\eeq
up to contact terms.\\
The l.h.s. of (\ref{37}) is given by\\
$$4\partial _{z_{1}}\partial _{\bar{z_{1}}}~\lim _{\beta _{1}\rightarrow 0}
\frac{\partial}{\partial \beta _{1}}~G_{3}(z_{1},z_{2},z_{3}\vert \beta _{1},
\beta _{2},\beta _{3})~.$$
Using (\ref{10}), (\ref{11}) the differentiation with respect to $\beta _{1}$
is straightforward. In the generic case $\beta _{2}\neq \beta _{3},~
\beta _{j}\neq 0,~~j=2,3$ one finds $A_{3}(0,\beta _{2},\beta _{3})=0$
Therefore,
the contribution
of terms with logarithms log$\vert z_{i}-z_{j}\vert $ generated by the
$\beta _{1}$ dependence of $\Delta _{1}$ drops out in the limit $\beta _{1}
\rightarrow 0$ and one obtains
after differentiation with respect to $z_{1}$
\beq
\langle \partial^{2}\phi (z_{1})e^{\beta _{2}\phi (z_{2})}
e^{\beta _{3}\phi (z_{3})}\rangle ~=~
\frac{4(\Delta _{2}-\Delta _{3})^{2}
\frac{\partial}{\partial \beta _{1}}
A_{3}(\beta _{1},\beta _{2},\beta _{3})
\vert _{\beta _{1}=0}}
{\vert z_{2}-z_{1}\vert ^{2(1+\Delta_{2}-\Delta_{3})}
\vert z_{1}-z_{3}\vert ^{2(1+\Delta_{3}-\Delta_{2})}
\vert z_{2}-z_{3}\vert ^{2(\Delta_{3}+\Delta_{2}-1)} }~
\label{39}
\eeq
up to contact terms that will be neglected.\\
This way
the quantum Liouville equation (\ref{37}) is reduced to
\beq
4(\Delta _{2}-\Delta _{3})^{2}
\frac{\partial}{\partial \beta _{1}}
A_{3}(\beta _{1},\beta _{2},\beta _{3})
\vert _{\beta _{1}=0}~=~\frac{\alpha \mu ^{2}}{2}~A_{3}(\alpha ,\beta _{2},
\beta _{3})~.
\label{40}
\eeq
This relation can easily be verified using the representation (\ref{14}) for
$A_3$ and the functional identities satisfied by the constituents $F_o,F_i$.
\\ \\ \\
{\Large \bf (iii) Liouville Two- and Three-Point Functions}\\ \\
%
%
%
%
%
As mentioned in the introduction and practised at an intermediate stage in the
previous section already, we relate the Liouville operator $\phi (z)$ to
$\partial _{\beta}e^{\beta \phi (z)}~=~\phi e^{\beta \phi (z)}$.
For reasons becoming clear in a moment
we still do not specify the value of $\beta $ after differentiation. We
only require to treat all $\beta _{j}$ on an equal footing.
{}From (\ref{11}), (\ref{10}), (\ref{6}) this yields
\bea
\partial _{\beta_{1}}\partial _{\beta_{2}}\partial _{\beta_{3}}G_{3}(
{}~z_{j}~\vert ~\beta _{j}~)&=&
\vert z_{1}-z_{2} \vert ^{2(\Delta _{3}-\Delta _{1}-\Delta _{2})}
\vert z_{1}-z_{3} \vert ^{2(\Delta _{2}-\Delta _{1}-\Delta _{3})}
\vert z_{2}-z_{3} \vert ^{2(\Delta _{1}-\Delta _{2}-\Delta _{3})}
\nonumber \\
\cdot &\Big (&\partial _{\beta_{1}}\partial _{\beta_{2}}\partial _{\beta_{3}}
A_{3}~
-~\sqrt {Q^{2}-8}~\partial _{\beta _{1}}\partial _{\beta _{2}}A_{3}(l_{12}~+~
l_{13}~+~l_{23})\nonumber \\
&+&(Q^{2}-8)~\partial _{\beta _{1}}A_{3}~(L_{13}~L_{12}~+~L_{23}~L_{12}~+~
L_{23}~L_{13})\hfill \nonumber \\
&+&(Q^{2}-8)^{\frac{3}{2}}A_{3}~L_{12}~L_{13}~L_{23}~\Big )~, \nonumber \\
\label{43}
\eea
with $l_{ij}~=~\mbox{log}\vert z_{i}-z_{j}\vert,~~~L_{ij}~=~
l_{ij}~-~l_{ki}~-~l_{kj},~~(i,j,k)~=~\mbox{perm}(1,2,3)$.\\

Use has been made of the equality of derivatives with respect to different
$\beta _{j}$ if after differentiation a symmetric point in $\beta $-space
is choosen.\\
The natural choice $\beta _{j}=0$ removes the power-like $z$-dependence
in (\ref{43}), but $A_{3}$ turns out to be singular at this point:
The value of $F_{1}F_{2}F_{3}$
depends on how the origin in $\beta $-space is approached. In addition
$F_{0}$ has a pole (note $s_{3}\rightarrow \frac{Q}{\alpha}=1+\frac{2}
{\alpha ^{2}}$, (\ref{24}), (\ref{25})).\\

Our functional integral yields the correlation
function of Liouville exponentials directly, there is no interpretation as
a vacuum expectation value with respect to a SL(2,C) invariant vacuum
\cite{Seiberg}. Therefore, operator insertions are well-defined only in the
presence of at least two
Liouville exponentials playing the role of spectators. This
concept worked perfectly in the previous section where we constructed
$\langle \phi (z_{1})e^{\beta _{2}\phi (z_{2})}e^{\beta _{3}\phi (z_{3})}
\rangle $. To get in the same sense 2 and 3-fold insertions of $\phi $
one has to start with the 4 and 5-point functions of exponentials.
Unfortunately,
these higher correlation functions are not available up to now.\\ \\ \\
{\Large \bf (iv) Poles and Zeros of $A_2$ and $A_3$}\\ \\
The spectrum of poles and zeros of the 3-point function and two special
degenerate cases of 4-point functions as well as related problems for the
interpretation of non-critical  strings have been discussed in \cite{DO2},
\cite{DO3}. We add in this section observations concerning the 3-point
and 2-point function which are relevant in connection with some
recent work on off shell critical strings \cite{MP} and which shed some
light on the question of mass shell conditions for non-critical strings.\\

For applications to noncritical strings we are interested in the case
Re $(\alpha ^{2})>0$. This of course is realized for $c_{M}<1$ but higher
dimensional target space $D>1$ made possible by the presence of a linear
dilaton background \cite{DO3} i.e.
\beq
c_{M}=D-3P^{2}.
\label{P}
\eeq
It is even valid
for $1\leq c_{M}<13$ if (\ref{6}) is taken seriously also in
between $1\leq c_{M}\leq 25$, since then we have
\beq
\alpha ^{2}~=~\frac{13~-~c_{M}~-~\sqrt{(25-c_{M})(1-c_{M})}}
{6}~.
\label{44}
\eeq
On the other side in ref. \cite{MP} off shell critical strings are constructed
for $c_{M}=26$ by enforcing the otherwise violated condition of
conformal (1,1) dimension by the dressing with suitable Liouville
exponentials. Clearly, for this application one needs $\alpha ^{2}<0$.\\
In the first situation Re $(\alpha ^{2})>0$ one finds
the following pole-zero pattern of $\prod _{j=1}^{3}F_{j}$ \cite{DO2}
\bea
\alpha ~\bar{\beta _{j}}&=&\frac{\alpha
^{2}}{2}~k_{j}~+~l_{j}~~~~(\mbox{poles})
\nonumber \\
\alpha ~\beta _{j}&=&\frac{\alpha ^{2}}{2}~k_{j}~+~l_{j}~~~~(\mbox{zeros})
\nonumber \\
\mbox{Re}~(\alpha^{2})~>0,~~~~~\mbox{integer}~k_{j},~l_{j},&&\mbox{both}~\leq
0~~~
\mbox{or both}~>0~.
\label{45}
\eea
While the position of zeros depends on the value of single $\beta _{j}$,
the pole position is given by a combination out of all $\beta _{j}$ involved.
Only in applications to dressings of minimal models also the pole position
factorizes (leg poles).\\
In the second situation Re $\alpha ^{2}<0$ one ends up with the following
situation
\bea
\alpha \beta _{j}~-~\frac{\alpha ^{2}}{2}&=&-~\frac{\alpha ^{2}}{2}~k_{j}~+
{}~l_{j}~~~~(\mbox{poles})
\nonumber \\
\alpha ~\bar{\beta _{j}}~-~\frac{\alpha ^{2}}{2}&=&-~\frac{\alpha ^{2}}{2}~
k_{j}~+~l_{j}~~~~(\mbox{zeros})
\nonumber \\
\mbox{Re}~(\alpha^{2})~<~0,~~~~~\mbox{integer}~k_{j},~l_{j},&&\mbox{both}~\leq
0~~~
\mbox{or both}~>0~.
\label{47}
\eea
Now the position of poles of $\prod _{j=1}^{3}F_{j}$ is determined by the
single
$\beta _{j}$.\\

The remaining factors in $A_{3}$ depend on the $\beta _{j}$
via $s_{3}$ only. For their combined pole-zero spectrum arising from the
$\Gamma$-functions and $F_{0}$ one finds for Re $\alpha ^{2} >0$ no zeros
but poles at
\bea
\frac{\alpha}{2}\sum _{i=1}^{3}\beta _{i}~-~\frac{\alpha ^{2}}{2}~-&1=&
\frac{\alpha ^{2}}{2}k~+~l~~~~(\mbox{poles})\nonumber \\
\mbox{Re}~(\alpha^{2})~>0,~~~~~\mbox{integer}~k,~l,&&\mbox{both}~\leq 0~~~
\mbox{or both}~>0~.
\label{b1}
\eea
In the other case Re $\alpha ^{2}<0$ one has instead
\bea
\frac{\alpha}{2}\sum _{i=1}^{3}\beta _{i}&=
&\frac{\alpha^{2}}{2}(1-j)~-~1~~~~(\mbox{poles})\nonumber \\
\frac{\alpha}{2}\sum _{i=1}^{3}\beta _{i}&=
&\frac{\alpha^{2}}{2}(k+2)~-~l~~~~(\mbox{zeros})\nonumber \\
\mbox{or}~~~~~
\frac{\alpha}{2}\sum _{i=1}^{3}\beta _{i}&=
&\frac{\alpha^{2}}{2}(1-k)~+~l+2~~~~(\mbox{zeros})\nonumber \\
\mbox{Re}~(\alpha^{2})~<0,~~~&&\mbox{integer}~k,~j,~l \geq 0~.
\label{b2}
\eea
Altogether we find a drastic
change in the analytic structure with respect to the $\beta _{j}$
in going from Re $\alpha ^{2} >0$ to Re $\alpha ^{2}<0$.\\

Let us turn to the 2-point function.We obtain
for arbitrary $\alpha ^{2}$
\bea
\alpha \beta &=&\frac{\alpha ^{2}}{2}~-~l~~\mbox{or}~~~
\alpha \beta~=~1~-~l~\frac{\alpha ^{2}}{2}~,~~\mbox{integer}~l~\geq ~0~~
\mbox{(poles)} \label{48} \\
\alpha \beta &=&\frac{\alpha ^{2}}{2}~+~j~~
\mbox{or}~~\alpha \beta~=~1~+~j~\frac{\alpha ^{2}}{2}~
{}~\mbox{or}~~\beta ~=~\frac{Q}{2}
,~~\mbox{integer}~j~
\geq ~2~~\mbox{(zeros).}
\nonumber
\eea

{}From this pole-zero pattern we can derive an interesting conjecture
concerning
the mass shell condition for noncritical strings. For instance the
coefficient $\beta $ in a gravitationally dressed vertex operator for tachyons
\cite{David,DK,DO1}
$$e^{ik_{\mu}X^{\mu}(z)}~e^{\beta \phi (z)}$$
is related to $k_{\mu}$ by the requirement of total conformal dimension (1,1)
(compare (\ref{e2}))
$$\frac{1}{2}\beta (Q-\beta )~+~\frac{k (k-P)}{2}~=~1~,$$
or equivalently
\beq
(\beta - \frac{Q}{2})^{2}~-~(k-\frac{P}{2})^{2}~=~\frac{1-D}{12}~.
\label{49}
\eeq
In contrast to the critical string, where the demand of dimension (1,1)
delivers the mass shell condition $\frac{k(k-P)}{2}=1$, eq. (\ref{49}) implies
no restriction for the target space momentum. \\

A condition on $k_{\mu}$
can arise only due to an additional restriction on the allowed values
of $\beta $.
The $\beta $ dependent factor $A_{2}$ discussed above appears as the dressing
factor in the 2-point S-matrix element for the tachyon excitation of the
string. From the point of view of field theory in target space this object is
an inverse propagator. Hence it should vanish as soon as the tachyon momentum
approaches its mass shell.
For generic $\beta $ the dressing factor $A_{2}(
\beta )$ is different from zero and it is natural to associate its zeros
with the mass shell. For $c_{M}<1$ i. e. $0<\alpha ^{2}<2$ the spectrum of
zeros
is unbounded from above. The lowest zero is $\beta =\frac{Q}{2}$.
The resulting spectrum for the mass of the gravitational dressed tachyon,
i.e. $m_{T}^{2}=\frac{1-D}{12}-(\beta -\frac{Q}{2})^{2}$, is not bounded
from below. However, since all zeros, except that at $\beta =\frac{Q}{2}$,
obey $\beta >\frac{Q}{2}$ they correspond to operators $e^{\beta \phi}$
describing states with wave functions in mini-superspace approximation
$\propto e^{(\beta -\frac{Q}{2})\phi}$. These states are not normalizable
in the infrared $\phi \rightarrow +\infty $ and have to be excluded \cite
{Seiberg}. On the other hand $\beta ~=~\frac{Q}{2}$ sits just on the border
to the ``microscopic" states describing local insertions with wave functions
peaked
in the ultraviolet and ``macroscopic" states with imaginary exponents.\\
$\beta ~=~Q/2$ then leads to
\beq
(k-\frac{P}{2})^{2}~=~\frac{D-1}{12}~.
\label{50}
\eeq
The generalization to higher string excitations is straightforward.
For instance in the graviton case an additional term $+1$ on the l.h.s.
of (\ref{49}) leads to $(k-\frac{P}{2})^{2}=\frac{D-25}{12}$.\\ \\
To conclude we contributed to the construction of correlation functions in
Liouville theory. This construction is a longstanding problem relevant
for various aspects of string theory and general conformal field theory.
We were able to calculate the 2 and 3-point functions of Liouville exponentials
of arbitrary real power. The method of continuation in the parameter $s$
passed a very crucial test. The Liouville equation of motion is fulfilled,
hence we are sure that the derived correlation functions indeed reflect
some essential features of quantized Liouville theory.
What concerns applications to noncritical string theory an interesting
conjecture on mass shell conditions emerged. Keeping the standard
picture that 2-point S-matrix elements vanish on shell, we
related the on shell condition to the spectrum of zeros of the 2-point
function of Liouville exponentials.
The further check of both the $s$-continuation itself as well as the spectrum
conjecture requires the knowledge of the higher ($N\geq 4$) correlation
functions. Unfortunately, at present the necessary integral formulas are
not available.


\end{document}